\documentclass[12pt]{article}
\usepackage{epsf}
\def\be{\begin{equation}}
\def\ee{\end{equation}}
\def\bea{\begin{eqnarray}}
\def\eea{\end{eqnarray}}

\def\ap#1#2#3 {Ann. Phys. (NY) {\bf#1} (19#2) #3}
\def\err#1#2#3 {{\it Erratum} {\bf#1} (19#2) #3}
\def\ib#1#2#3 {{\it ibid.} {\bf#1} (19#2) #3}
\def\ijmp#1#2#3 {Int. J. Mod. Phys. {\bf#1} (19#2) #3}
\def\jetp#1#2#3 {JETP Lett. {\bf#1} (19#2) #3}
\def\mpl#1#2#3 {Mod. Phys. Lett. {\bf#1} (19#2) #3}
\def\np#1#2#3 {Nucl. Phys. {\bf#1} (19#2) #3}
\def\pl#1#2#3 {Phys. Lett. {\bf#1} (19#2) #3}
\def\prep#1#2#3 {Phys. Rep. {\bf#1} (19#2) #3}
\def\prev#1#2#3 {Phys. Rev. {\bf#1} (19#2) #3}
\def\prl#1#2#3 {Phys. Rev. Lett. {\bf#1} (19#2) #3}
\def\sjnp#1#2#3 {Sov. J. Nucl. Phys. {\bf#1} (19#2) #3}
\def\spj#1#2#3 {Sov. Phys. JETP {\bf#1} (19#2) #3}
\def\spu#1#2#3 {Sov. Phys. Usp. {\bf#1} (19#2) #3}
\def\zp#1#2#3 {Zeit. Phys. {\bf#1} (19#2) #3}

\begin{document}
\begin{titlepage}
\begin{center}
{\Large \bf William I. Fine Theoretical Physics Institute \\
University of Minnesota \\}
\end{center}
\vspace{0.2in}
\begin{flushright}
FTPI-MINN-19/27 \\
UMN-TH-3904/19 \\
November 2019 \\
\end{flushright}
\vspace{0.3in}
\begin{center}
{\Large \bf Update on splitting of lifetimes of $c$ and $b$ hyperons within the heavy quark expansion and decays $\Xi_Q \to \Lambda_Q \pi$ 
\\}
\vspace{0.2in}
{\bf  M.B. Voloshin  \\ }
William I. Fine Theoretical Physics Institute, University of
Minnesota,\\ Minneapolis, MN 55455, USA \\
School of Physics and Astronomy, University of Minnesota, Minneapolis, MN 55455, USA \\ and \\
Institute of Theoretical and Experimental Physics, Moscow, 117218, Russia
\\[0.2in]

\end{center}

\vspace{0.2in}

\begin{abstract}
It is argued that the results of recent precision measurements by LHCb of the lifetimes of charmed and bottom hyperons are very well consistent with the description within heavy quark expansion and allow to accurately determine  matrix elements of light-flavor nonsinglet four-quark operators over the hyperons and to accurately reproduce the difference of lifemes of $\Lambda_b$ and $\Xi_b^-$. When combined with the recent LHCb results on the decay $\Xi_b^- \to \Lambda_b \pi^-$ this leads to prediction of a lower bound on the rate of the decays $\Xi_c \to \Lambda_c \pi$. 
  \end{abstract}
\end{titlepage}

Experimental and theoretical studies of the differences of inclusive weak decay rates of hadrons containing a heavy quark ($c$ or $b$) attract considerable interest ever since the first observation~\cite{delco} of unequal lifetimes of charged and netral charmed $D$ mesons. These differences, substantial among the charmed hadrons and significantly smaller for the bottom ones, are due to the light quark/qluon degrees of freedom in a hadron with heavy quark $Q$, and are suppressed by inverse powers of the heavy mass $m_Q$. Thus the inclusive decay rates of the discussed hadrons are described~\cite{sv0,bgt,vs,vs1,buv}  in terms of a systematic heavy quark expansion in powers of $m_Q^{-1}$. (A recent review of the theoretical development can be found in Ref.~\cite{Lenz}.) The terms of this expansion contain quark/gluon operators of appropriate dimension, whose matrix elements over a hadron describe the contributions to the inclusive weak decay rates of that meson or hyperon. Theoretical evaluations of these matrix elements are highly model dependent. However, within an application of the heavy quark expansion to both charmed and bottom hadrons, the matrix elements do not depend on the heavy quark flavor, giving rise to relations between the differences of the inclusive decay rates in the $c$ and $b$ sectors. Such relations for heavy baryons~\cite{mv99,mv00} were for a long time in contradiction with the measurements of the ratio of the lifetimes $\tau(\Lambda_b)/\tau(B_d)$ by the LEP experiments~\cite{aleph1,aleph2,aleph3,opal}, with the experimental value of the ratio being too low (around 0.8), which was also at variance with the original prediction~\cite{vs1} (0.95 - 1). The situation has changed dramatically with the greatly improved the precision of the measurement by LHCb of $\tau(\Lambda_b)$~\cite{lhcb1,lhcb2}, invalidating the old CERN results and leading to the  
current average value~\cite{pdg} $\tau(\Lambda_b)/\tau(B_d)=0.964 \pm 0.007$ very well consistent with the theory expectations.

It should be mentioned that the latter ratio is sensitive to several terms in the expansions, namely the contribution of the four-quark operators~\cite{vs1} and of the chromo-magnetic term~\cite{buv}. Thus the relation between the decay rates of charmed and bottom hadrons is still somewhat model-dependent~\cite{mv00}. The theoretical uncertainty is greatly reduced in the relation between the differences of the lifetimes in the flavor $SU(3)_f$ (anti)triplet of charmed hyperons ($\Lambda_c, \, \Xi_c^+$ and $\Xi_c^0$) and similar differences for the $b-$hyperons  ($\Lambda_b, \, \Xi_b^0$ and $\Xi_b^-$). The splittings of the inclusive decay rates in these antitriplets are expressed~\cite{mv99} in terms of two differences of diagonal matrix elements of four-quark operators: 
differences of diagonal
matrix elements over the hyperons:
\begin{eqnarray}
\label{defxy}
&& \! \! \! \! \! \! \! \!x=\left \langle  {1 \over 2} \, (\overline Q \, \gamma_\mu \, Q) \left
[
(\overline u \, \gamma_\mu u) - (\overline s \, \gamma_\mu s) \right]
\right \rangle_{\Xi_{Qd}-\Lambda_Q} = \left \langle  {1 \over 2} \,
(\overline Q \, \gamma_\mu \, Q) \left [ (\overline s \, \gamma_\mu s) -
(\overline d \, \gamma_\mu d) \right] \right \rangle_{\Lambda_Q -
\Xi_{Qu}}~,  \\ \nonumber
&& \! \! \! \! \! \! \! \! \! \! y=\left \langle  {1 \over 2} \, (\overline Q_i \, \gamma_\mu \, Q_k)
\left [ (\overline u_k \, \gamma_\mu u_i) - (\overline s_k \, \gamma_\mu
s_i) \right ] \right \rangle_{\Xi_{Qd}-\Lambda_Q} = \left \langle  {1
\over 2} \, (\overline Q_i \, \gamma_\mu \, Q_k) \left [ (\overline s_k
\, \gamma_\mu s_i) - (\overline d_k \, \gamma_\mu d_i) \right] \right
\rangle_{\Lambda_Q - \Xi_{Qd}}
\end{eqnarray}
with the notation for the differences of the matrix elements:
$\langle {\cal O} \rangle_{A-B}= \langle A | {\cal O} | A \rangle -
\langle B | {\cal O} | B \rangle$. Also in these expressions $Q$ stands for the heavy $c$ or $b$ quark, and the corresponding particle in the antitriplet of hyperons are denoted as $\Lambda_Q \sim Qud,\, \Xi_{Qu} \sim Qsu, \, \Xi_{Qd} \sim Qsd$. Finally the indices $i,\,k$ label the color of quarks, and the nonrelativistic normalization of the heavy quark operators, $\langle Q| Q^\dagger Q |Q \rangle=1$ is assumed throughout this paper. 

The matrix elements in Eq.(\ref{defxy}) do not depend on the mass of the heavy quark (provided that the operators are normalized at a fixed low scale $\mu$ so that the heavy quark can be considered as a static source). Thus one can relate the differences of inclusive decay rates of the charmed and bottom hyperons. The present paper is triggered by the most recent LHCb precision measurement~\cite{lhcb19} of the lifetimes of the charmed hyperons that results in a very significant modification of estimates, as compared to the ones~\cite{mv99,mv00} based on earlier data. The most significant modification originates from the shift of the measured lifetime of $\Xi_c^0$ from  $(112^{+13}_{-10}) \,$fs (still the current PDG value~\cite{pdg}) to the recent LHCb result $(154.5 \pm 1.7 \pm 1.6 \pm 1.0)\,$fs. In particular an analysis using old data resulted in a predicted~\cite{mv99} difference of inclusive decay rates of the $\Lambda_b$ and $\Xi_b^-$, $\Delta_b = (0.11 \pm 0.03)\,$ps$^{-1}$, whereas with the new data it is found here to be $(67 \pm 2)\times 10^{-3} \,$ps$^{-1}$, which puts it within the range of experimental errors and theoretical uncertainties from the current data on the lifetimes of $\Lambda_b$ and $\Xi_b^-$.

Furthermore, the same matrix elements enter a current algebra relation~\cite{mv00-2} for the difference between the $S$-wave amplitudes of the strangeness decay in the heavy hyperons, $\Xi_c \to \Lambda_c \pi$ and $\Xi_b \to \Lambda_b \pi$. Therefore the latter difference can also be evaluated from measured lifetime splittings of the charmed hyperons.  It will be argued here that with the new data the difference is substantially smaller than the amplitude corresponding to the recently measured\cite{lhcb15,lhcb19-0} rate of the decay $\Xi_b^- \to \Lambda_b \pi^-$. Thus it is possible to estimate the lower bound on the rate of the charmed hyperon decay $\Xi_c \to \Lambda_c \pi$. 

As is already mentioned, the variations in the inclusive weak decay rates of hadrons with the same heavy flavor are calculated in terms of expansion in inverse powers of the heavy quark mass. These variations, corresponding to different flavor of the spectator light quark, appear as  terms of order $m_Q^{-3}$ (in comparison with the leading `parton' decay rate of the heavy quark proportional to $m_Q^5$, and are described by matrix elements of four-quark operators over the hadrons, see e.g in Ref.~\cite{Lenz}). At $\mu \ll m_Q$ there arises a so called `hybrid' ~\cite{vs1, vs2} QCD renormalization of the operators from the
normalization scale $m_Q$ down to a low scale $\mu$ depending on the parameter $\alpha_s(\mu)/\alpha_s(m_Q)$. The full formulas for the discussed decay rate differences between the heavy hyperons can be found in Ref.~\cite{mv99}. The expressions in the charm sector somewhat simplify with the choice of $\mu =m_c$ which is assumed throughout this paper. The terms arising at the four-quark operator level in the expansion for the charmed  hadrons are expressed through six coefficiens $C_1, \ldots, C_6$:
\begin{eqnarray}
&&C_1= C_+^2+C_-^2 ~,
\nonumber \\
&&C_2=  C_+^2-C_-^2~, \nonumber \\
&&C_3=- {1 \over 4} \, (C_+-C_-)^2 ~, \nonumber \\
&&C_4=-{1 \over 4} \,  (5C_+^2+C_-^2+6C_+C_-)~, \nonumber
\\
&&C_5=-{1 \over 4} \,  (C_++C_-)^2 ~, \nonumber \\
&&C_6=-{1 \over 4}  \, (5C_+^2+C_-^2-6C_+C_-)~,
\label{coefs}
\end{eqnarray}
with $C_+$ and $C_-$ being the
standard coefficients in the QCD renormalization of the non-leptonic
weak interaction from $m_W$ down to the charmed quark mass:
$C_-=C_+^{-2}=(\alpha_s(m_c)/\alpha_s(m_W))^{4/b}$, where $b$, the
coefficient in the one-loop beta function in QCD, can be taken as
$b=25/3$ for the case of the charmed quark decay.
The discussed here calculation of the differences of the lifetimes for the charmed hyperons takes into account the CKM dominant as well as single Cabibbo suppressed nonleptonic and semileptonic decays. The relevant parts of the effective Lagrangian whose average over a charmed hadron gives the correction to the corresponding inclusive decay rate is written in terms of $C_A$ as~\cite{vs1,mv99}
\begin{eqnarray}
&&L_{nl,0}=  c^4 \,{G_F^2 \, m_c^2 \over 4 \pi} \,
\left [
C_1 \, (\overline c \Gamma_\mu c)(\overline d \Gamma_\mu d) + C_2  \,
(\overline c \Gamma_\mu d) (\overline d \Gamma_\mu c) +\right .
\nonumber \\
&& C_3 \, (\overline  c \Gamma_\mu c +
{2 \over 3}\overline c \gamma_\mu \gamma_5 c) (\overline s \Gamma_\mu
s)+ C_4 \, (\overline  c_i \Gamma_\mu c_k +
{2 \over 3}\overline c_i \gamma_\mu \gamma_5 c_k)
(\overline s_k \Gamma_\mu s_i) +
\nonumber \\
&& \left . C_5 \, (\overline  c \Gamma_\mu c +
{2 \over 3}\overline c \gamma_\mu \gamma_5 c) (\overline u \Gamma_\mu
u)+ C_6 \, (\overline  c_i \Gamma_\mu c_k +
{2 \over 3}\overline c_i \gamma_\mu \gamma_5 c_k)
(\overline u_k \Gamma_\mu u_i)
\right ]~,
\label{nl0}
\end{eqnarray}

\begin{eqnarray}
&&L_{nl,1}= c^2 \, s^2 \,{G_F^2 \, m_c^2 \over 4 \pi} \,
\left [
C_1 \, (\overline c \Gamma_\mu c)(\overline q \Gamma_\mu q) + C_2  \,
(\overline c_i \Gamma_\mu c_k) (\overline q_k \Gamma_\mu q_i) +\right .
\nonumber \\
&& C_3 \, (\overline  c \Gamma_\mu c +
{2 \over 3}\overline c \gamma_\mu \gamma_5 c) (\overline q \Gamma_\mu
q)+ C_4 \, (\overline  c_i \Gamma_\mu c_k +
{2 \over 3}\overline c_i \gamma_\mu \gamma_5 c_k)
(\overline q_k \Gamma_\mu q_i) +  \nonumber \\
&& \left . 2 \, C_5 \, (\overline  c \Gamma_\mu c +
{2 \over 3}\overline c \gamma_\mu \gamma_5 c) (\overline u \Gamma_\mu
u)+ 2 \, C_6 \, (\overline  c_i \Gamma_\mu c_k +
{2 \over 3}\overline c_i \gamma_\mu \gamma_5 c_k)
(\overline u_k \Gamma_\mu u_i)
\right ]~,
\label{nl1}
\end{eqnarray}

\be
L_{sl} = - {G_F^2 \, m_c^2 \over 2 \pi} \, \left [  c^2 \, (\overline  c_i \Gamma_\mu c_k +  {2 \over 3}\overline c_i
\gamma_\mu \gamma_5 c_k)
(\overline s_k \Gamma_\mu s_i) +s^2 \, (\overline  c_i \Gamma_\mu c_k +
{2 \over 3}\overline c_i \gamma_\mu \gamma_5 c_k)
(\overline d_k \Gamma_\mu d_i) \right ]~,
\label{sl}
\ee
where $c = \cos \theta_c$ $s=\sin \theta_c$, $\Gamma_\mu = \gamma_\mu (1-\gamma_5)$, and the the notation $(\overline q \, \Gamma \, q)= (\overline d \, \Gamma
\, d) + (\overline s \, \Gamma \, s)$ is used. The subscript in the notation for the effective Lagrangian in Eqs. (\ref{nl0}) and (\ref{nl1}) indicates the order of the Cabibbo suppression, and the overall coefficient in Eq.(\ref{sl}) takes into account two inclusive semileptonic channels, with $e \nu$ and with $\mu \nu$.

Averaging of the effective Lagrangian over charmed hyperons is greatly simplified by their spin structure. Namely, the light quark pair has total spin 0 and there is no correlation between the spin of any of the light quarks with that of the nonrelativistic heavy quarks. Therefore it is only the Vector$\times$Vector part of the four-quark operators that has a nonzero average. Using also the flavor $SU(3)$ symmetry one can use the formulas (\ref{nl0}) - (\ref{sl}) to express the differences in (semi)inclusive decay rates of the baryons in terms of two parameters $x$ and $y$ defined in Eq.(\ref{defxy})~\footnote{The euality in each line of Eq.(\ref{defxy}) of the two expressions for $x$ (and separately for $y$) is obviously guaranteed by the isospin symmetry. However a derivation of the differences of the averages of the operators in terms of $x$ and $y$ has to rely on the flavor $SU(3)$ symmetry.}~. The semi-inclusive decay rates for charmed baryons are not yet well known. Thus one has to use the data on the total decays rates, i.e. on the lifetimes, that are reasonably well measured by now~\cite{lhcb19}. Using the equations (\ref{nl0}) - (\ref{sl}) and also (\ref{coefs}) one arrives at the following expressions for the diefferences of the total decay rates in the antitriplet of charmed hyperons
\bea
&&\Delta_1 \equiv \Gamma (\Xi_c^0) - \Gamma (\Lambda_c) = - c^2 \, {G_F^2 \, m_c^2 \over 16 \pi} \, \left \{ \, \left [ 4  c^2 C_- C_+  + s^2 ( 5 C_-^2 + 5 C_+^2 + 6 C_- C_+) \right ] \, x - \right . \nonumber \\
&&\left . \left [ 8+12 c^2 C_- C_+ + 3 s^2 \, (3 C_-^2 + 18 C_-C_+ - 9 C_+^2) \right ] \, y \right \}~,
\label{Delta1}
\eea

\bea
&&\Delta_2 = \Gamma
(\Lambda_c) - \Gamma (\Xi_c^+) = - {G_F^2 \, m_c^2 \over 4 \pi} \, \left \{  c^4 \left [ C_-^2 +  C_+^2 + {1 \over 4} \, (C_+-C_-)^2 \right ] \, x + \right .  \nonumber \\
&&\left . c^4 \left [ C_+^2 - C_-^2 +{1 \over 4} \, (C_-^2+5 C_+^2+6 C_+C_-) \right] \, y + 2 (c^2-s^2) \, y\right \}~.
\label{Delta2}
\eea

Using a realistic value for the QCD coupling $\alpha_s(m_c)/\alpha_s(m_W) \approx 2.5$, one finds\footnote{The final results rather weakly depend on this numeric assumption.} $C_- \approx 1.55$ and $C_+ \approx 0.8$, and the relations (\ref{Delta1}) and (\ref{Delta2}) read numerically
\bea
&& \left ( {\Delta_1 \over {\rm ps}^{-1}} \right ) \, \left ( {1.4 {\rm GeV} \over m_c } \right )^2 =-44.86 \,  \left (  {x \over {\rm GeV}^3} \right ) + 178.8 \,  \left (  {y \over {\rm GeV}^3} \right )~, \nonumber \\
&&\left ( {\Delta_2 \over {\rm ps}^{-1}}  \right ) \, \left ( {1.4 {\rm GeV} \over m_c } 
\right )^2 =-92.75 \, \left ( {x \over {\rm GeV}^3} \right ) - 101.8 \, \left ( {y \over {\rm GeV}^3} \right )~.
\label{numdelta}
\eea

The reported~\cite{lhcb19} by LHCb lifetimes of the charmed baryons correspond to the total decay rates 
$\Gamma(\Lambda_c^+) = (4.866 \pm 0.024 \pm 0.031 \pm 0.033)\,$ps$^{-1}$, $\Gamma(\Xi_c^+) = (2.189 \pm 0.017 \pm 0.014 \pm 0.015)\,$ps$^{-1}$ and $\Gamma(\Xi_c^0) = (6.472 \pm 0.071 \pm 0.067 \pm 0.041)\,$ps$^{-1}$, where the last error is in a common normalization factor due to the uncertainty in the $D^+$ lifetime.  Using these values and Eq.(\ref{numdelta}) one readily finds the numerical values of the parameters $x$ and $y$:
\bea
&&\left (  {x \over {\rm GeV}^3} \right ) = -(30.4 \pm 0.5 \pm 0.2) \times 10^{-3} \, \left ( {1.4 {\rm GeV} \over m_c } \right )^2 , \nonumber \\ 
&&\left (  {y \over {\rm GeV}^3} \right ) = (1.4 \pm 0.5 \pm 0.2) \times 10^{-3} \, \left ( {1.4 {\rm GeV} \over m_c } \right )^2~, 
\label{nxy}
\eea
where the last error is from the overall normalization of the data and the rest of the error is a result of addition in quadrature of the statistical and systematic experimental~\cite{lhcb19} errors. It can be reminded that the parameter $x$ does not depend on the scale $\mu$ below $m_c$, while the parameter $y$ depends on the normalization point and the value above corresponds to $\mu=m_c$. The numerical results in Eq.(\ref{nxy}) differ from and have much smaller ``experimental" errors than those found in Ref.~\cite{mv00} using the old data. 

The results (\ref{nxy}) for the matrix elements (\ref{defxy}) can be further used for evaluating the lifetime differences among the $b$ hyperons. Taking into account only the dominant $b \to c$ weak interaction transition, one readily concludes that the latter differences are contributed only by the nonleptonic decays, and also that the effective Lagrangian for spectator quark effects is symmetric with respect to $s \leftrightarrow d$, i.e. it corresponds to $\Delta U =0$~\footnote{This property is broken by the small kinematical effects of the $c$ quark mass in the effective Lagrangian. Phenomenologically the smallness of the $U$ symmetry breaking in the discussed correction is known from very small difference of the decay rates between $B_d$ and $B_s$ mesons.} and thus there is no splitting between the decay rates of $\Lambda_b$ and $\Xi_b^0$: $\Gamma(\Lambda_b)= \Gamma(\Xi_b^0)$. The remaining nonzero rate difference is expressed in terms of $x$ and $y$ as~\cite{mv99}
\bea
&&\Delta_b \equiv \Gamma(\Lambda_b) - \Gamma(\Xi_b^-) = - c^2 \, |V_{bc}|^2 
\,{G_F^2 \, m_b^2 \over 16 \pi} \, \times \nonumber \\
&&\left \{ \left [ (4+\xi) \,  \tilde C_-^2 + (8-3\, \xi) \tilde C_+^2 + 2 \xi \, \tilde C_- \tilde C_+ \right ] \, x + 3 \xi \, (3 \tilde C_+^2 - \tilde C_-^2 - 2 \tilde C_- \tilde C_+) y \right \}~,
\label{delb}
\eea
where the renormalization coefficients are determined by $\alpha_s(m_b)$: $\tilde C_- = \tilde C_+^{-2} = [\alpha_s(m_b)/\alpha_s(m_W)]^{4/b}$ and the coefficient $\xi = [\alpha_s(m_c)/\alpha_s(m_b)]^{1/2}$ describes the hybrid renormalization of the four-quark orepators below $m_b$ down to $\mu = m_c$. (A realistic value $\xi \approx 1.12$ is used here. The numerical results only weakly depend on this parameter.) The decay rate difference (\ref{delb}) can be evaluated using the results (\ref{nxy}). The estimated errors in the parameters $x$ and $y$ are however strongly correlated, and it might be more convenient to use these parameters as solutions to Eqs.~(\ref{Delta1}) and (\ref{Delta2}) in terms of the rate splittings $\Delta_1$ and $\Delta_2$. Proceeding in this way one finds~\cite{mv99}
\bea
&&\Delta_b = |V_{bc}|^2  \, {m_b^2 \over m_c^2} \, (0.85 \,
\Delta_1 + 0.91 \, \Delta_2) \approx (15 \, \Delta_1 + 16 \,
\Delta_2) \times 10^{-3} = \nonumber \\
&&\left \{ 15 \, [\Gamma(\Xi_c^0) - \Gamma(\Xi_c^+)] +   [\Gamma(\Lambda_c^+) - \Gamma(\Xi_c^+)] \right \} \times 10^{-3} = (67 \pm 2) \times 10^{-3} \, {\rm ps}^{-1}~.
\label{dbres}
\eea
It is clear from this expression that this estimate of the expected decay rate difference for the $b$ hyperons is
mostly sensitive to the input for $\Gamma(\Xi_c^0) - \Gamma(\Xi_c^+)$ (and has only very little sensitivity to $\Gamma(\Lambda_c^+)$). In new LHCb data~\cite{lhcb19} this difference is greatly reduced, resulting  in 
the final numerical value in Eq.(\ref{dbres}) being  almost two times smaller than with the previous data in Ref.~\cite{mv99}. The indicated error, resulting from the experimental uncertainties is even more strongly reduced, so that the overall uncertainty in $\Delta_b$ is certainly dominated by the theoretical approximations and assumptions. 

It is quite satisfying to note that the estimate (\ref{dbres}) is in a greatly improved agreement with the data. Indeed, using the value of the $\Lambda_b$ lifetime from the Tables~\cite{pdg}, $\tau(\Lambda_b) = (1.471 \pm 0.009)\,$ps, and the estimated value of $\Delta_b$ one gets for the central value of $\Gamma^{-1}(\Xi_b^-)$ the numerical estimate 1.63\,ps, which is only $1.5 \, \sigma$ away from the experimental average $(1.57 \pm 0.04)\,$ps. 

The agreement is further slightly improved if one takes into account the decay of strangeness in $\Xi_b^-$:  $\Xi_b^- \to \Lambda_b \pi^-$, which is not included in the presented counting of the $b$ quark decay effects, and whose branching fraction is indicated~\cite{lhcb15,lhcb19-0} to be at the level of one percent. Besides their contribution to the overall balance of the lifetimes of the heavy hyperons, the decays of this type are of an interest on their own~\cite{mv00-2,Cheng92,sk,lv,fm,gr1,gr2,Cheng15}. The strangeness decay in the $b$ hyperons is induced by the underlying `spectator' decay of the strange quark, $s \to u \bar u d$. In the charmed hadrons, in addition to the spectator decay, there is a contribution of `non-spectator' weak scattering, $sc \to cd$. It is thus quite natural that the difference between the amplitudes of $\Xi_c \to \Lambda_c \pi$ and $\Xi_b \to \Lambda_b \pi$ is related to the same lifetime differences. Namely, in the heavy quark limit the spectator emission of a pion is a static $0^+ \to 0^+$ transition and thus proceeds only in the $S$ wave~\cite{mv00-2}. The non-spectator part generally gives rise to both $S$ and $P$ wave emission of a pion. The $S$ wave amplitude $A_S$ is not vanishing at zero momentum of the pion and can be evaluated using the PCAC relation
\be
\langle \Lambda_Q \, \pi_i (p=0) \,| H_W |\, \Xi_Q \rangle = {\sqrt{2}
\over f_\pi} \, \langle \Lambda_Q \, |\left[Q^5_i, \, H_W \right ] |\,
\Xi_Q \rangle~,
\label{pcac}
\ee
with $Q^5_i$ being the isotopic axial charges, $f_\pi \approx 130\,$MeV the pion decay constant,  and the weak Hamiltonian (normalized at $\mu = m_c$) describing both the spectator and non-spectator decay of strangess having the standard form
\bea
H_W = &&\sqrt{2} \, G_F \, c\, s \, \left \{ \left ( C_+ + C_-
\right ) \, \left [ (\overline u_L \, \gamma_\mu \, s_L)\, (\overline
d_L \, \gamma_\mu \, u_L) - (\overline c_L \, \gamma_\mu \, s_L)\,
(\overline d_L \, \gamma_\mu \, c_L) \right ] + \right . \nonumber \\
&& \left. \left ( C_+ - C_- \right ) \, \left [ (\overline d_L \,
\gamma_\mu \, s_L)\, (\overline u_L \, \gamma_\mu \, u_L) - (\overline
d_L \, \gamma_\mu \, s_L)\, (\overline c_L \, \gamma_\mu \, c_L) \right
] \right \}~.
\label{hw}
\eea

The spectator part of the amplitude is the same in the decays of charmed and bottom strange baryons while the non-spectator one gives an extra contribution to the $S$-wave decays of $\Xi_c$. Using the equations (\ref{pcac}) and (\ref{hw}) one arrives at the relation~\cite{mv00-2}
\bea
&&\Delta A_S \equiv \langle \Lambda_c \, \pi^- (p=0) \,| H_W |\, \Xi_c^0 \rangle - \langle
\Lambda_b \, \pi^- (p=0) \,| H_W |\, \Xi_b^- \rangle =  \nonumber \\
&&{\sqrt{2} \over f_\pi} \, G_F \, c \, s \, \langle \Lambda_c \,
| \left ( C_+ + C_- \right ) \, (\overline c_L \, \gamma_\mu \, s_L)\,
(\overline u_L \, \gamma_\mu \, c_L) + \nonumber \\
&&\left ( C_+ - C_- \right ) \, (\overline u_L \, \gamma_\mu \, s_L)\,
(\overline c_L \, \gamma_\mu \, c_L) |\, \Xi_c^0 \rangle = \nonumber \\
&&{G_F \, c\, s \over 2 \, \sqrt{2} \, f_\pi} \, \left[ \left ( C_- -
C_+ \right ) \, x - \left ( C_+ + C_- \right ) \, y \right ] ~,
\label{xic}
\eea
where $x$ and $y$ are the same matrix elements as in Eq.(\ref{defxy}) and the last transition makes use of the $SU(3)$ symmetry to relate the matrix elements between $\Xi_c$ and $\Lambda_c$ to the difference of diagonal averages. (The considered processes are pure $\Delta I = 1/2$ in the heavy quark limit~\cite{mv00-2}, so that the rates of the decays with emission of $\pi^0$, $\Xi_c^0 \to \Lambda_c^+ \pi^0$ and $\Xi_b^0 \to \Lambda_b \pi^0$ are simply half those for the emission of $\pi^-$.) When expressed in terms of the previously introduced total decay rate differences $\Delta_1$ and $\Delta_2$ the last formula in Eq.(\ref{xic}) reads as~\cite{mv00-2}
\bea
&& \Delta A_S \approx - {\sqrt{2} \, \pi \, c \, s \over G_F \,
m_c^2 \, f_\pi} \, \left (  0.45 \, \Delta_1 + 0.04 \Delta_2 \right ) = \nonumber \\
&&-10^{-7} \left \{
0.97 \left [ \Gamma(\Xi_c^0)-\Gamma(\Lambda_c) \right ] + 0.09 \left [
\Gamma(\Lambda_c)-\Gamma(\Xi_c^+) \right ] \right \} \, \left ( {1.4 \,
GeV \over m_c} \right )^2 \, {\rm ps} \nonumber \\ &&\approx 
-(1.8 \pm 0.1) \times 10^{-7}~.
\label{dasm}
\eea
This formula shows that the quantity $\Delta A_S$ is mostly determined by the difference of total decay rates $\Gamma(\Xi_c^0)-\Gamma(\Lambda_c)$. In the new data~\cite{lhcb19} this difference is approximately three times smaller than its old value. Hence the final estimate is proportionally smaller in comparison with the previous evaluations~\cite{mv00-2,gr2} based on old data.

The $S$ wave contribution to the decay rate is given in terms of $A_S$ as
\be
\Gamma_S (\Xi_Q - \Lambda_Q \pi) = |A_S|^2 \, {p_\pi \over 2 \pi}~,
\ee
where $p_\pi$ is the momentum of the emitted pion. Thus the rate that would correspond to the difference $\Delta A_S$ estimated in Eq.(\ref{dasm}) is approximately $0.9 \times 10^{-3}\,$ps$^{-1}$ with an error that is much smaller than other uncertainties. By combining the LHCb experimental results reported in Ref.~\cite{lhcb15} and Ref.~\cite{lhcb19-0} one can rather approximately estimate the branching fraction ${\cal B}(\Xi_b^- \to \Lambda_b \pi^-) \approx (0.8 \pm 0.3) \times 10^{-2}$, corresponding to the decay rate $\Gamma(\Xi_b^- \to \Lambda_b \pi^-) \approx (5 \pm 2) \times 10^{-3}$ps$^{-1}$. Clearly, this value is more than two sigma above the estimated rate corresponding to $\Delta A_S$. Thus even under least favorable assumption the of destructive interference between the spectator and non-spectator amplitudes the $S$-wave amplitude for the decay $\Xi_c^0 \to \Lambda_c^+ \pi^-$ is likely nonzero with the `formal' estimate
$\Gamma(\Xi_c^0 \to \Lambda_c^+ \pi^-) > \Gamma_{\rm min}$ with $\Gamma_{min} \approx (1.6  \pm 1.0) \times 10^{-3}\,$ps$^{-1}$, corresponding to ${\cal B}(\Xi_c^0 \to \Lambda_c^+ \pi^-) > {\cal B}_{\rm min} \approx (0.25 \pm 0.15) \times 10^{-3}$. Certainly in the case of constructive interference the lower bound for the decay rate of the $\Xi_c^0$ baryon is just slightly larger than the rate of the bottom hyperon decay decay $\Xi_b^- \to \Lambda_b \pi^-$. 

It should be emphasized that the equation (\ref{xic}) contains only the $S$-wave part of the non-spectator amplitude. The $sc \to c d$ scattering however can produce a pion in the $P$ wave. For this reason an estimate based on this equation can give only the lower bound for the decay rate  $\Gamma(\Xi_c^0 \to \Lambda_c^+ \pi^-)$ but not the full rate.

In summary. The splittings of lifetimes of heavy baryons are described by the heavy quark expansion for the inclusive decay rates. An updated analysis based on recent precision measurements of the lifetimes of charmed hyperons greatly im proves the evaluation of the relevant hadronic matrix elements of the four-quark operators [Eq.(\ref{defxy})]. The calculated contribution of the four-quark term in the expansion to the difference of the lifetimes of the $b$ barions, $\Xi_b^-$ and $\Lambda_b$, agrees with the data within about 1.5\,$\sigma$. Thus at the present level of accuracy this term is sufficient for description of the lifetime splittings among the charmed and bottom hadrons and no effects of higher terms show up, even though in the charmed sector the contribution of the four-quark term is large. Furthermore, the same matrix elements (\ref{defxy}) determine the difference of the $S$-wave amplitudes of the decays of strangeness $\Xi_c \to \Lambda_c \pi$ and $\Xi_b \to \Lambda_b \pi$,  which difference is due to the weak scattering $sc \to cd$ that is present only in the charmed hadrons.  An evaluation with the new data gives a significantly smaller estimate of this difference that is also substantially smaller than the central value of the amplitude for the recently observed decay $\Xi_b^- \to \Lambda_b \pi^-$ and implies a nonzero lower bound for the rate of the decays $\Xi_c \to \Lambda_c \pi$. The latter analysis would greatly benefit from even a modest improvement of the accuracy of the data on $\Xi_b^- \to \Lambda_b \pi^-$.

This work is supported in part by U.S. Department of Energy Grant No.\ DE-SC0011842.


\begin{thebibliography}{99}
\bibitem{delco} 
  W.~Bacino {\it et al.},
  %``Evidence for Unequal Lifetimes of the D0 and D+,''
  Phys.\ Rev.\ Lett.\  {\bf 45}, 329 (1980).
\bibitem{sv0}
M.~A.~Shifman and M.~B.~Voloshin, (1981) unpublished, presented in the
review V.~A.~Khoze and M.~A.~Shifman, Sov.\ Phys.\ Usp.\  {\bf26}, {387} (1983).
\bibitem{bgt}
N.~Bilic, B.~Guberina and J.~Trampetic, Nucl.\ Phys.\ B {\bf 248}, {261} (1984).
\bibitem{vs}
M.~A.~Shifman and M.~B.~Voloshin, Sov.\ J.\ Nucl.\ Phys.\ {\bf 41}, {120} (1984).
\bibitem{vs1}
M.~A.~Shifman and M.~B.~Voloshin, Sov.\ Phys.\ JETP {\bf 64}, {698} (1986).
\bibitem{buv}
I.~I.~Bigi, N.~G.~Uraltsev, and A.~I.~Vainshtein, Phys.\ Lett.\ B {\bf 293}, {430} (1992) , erratum -- ibid. {\bf 297}, 477 (1993).
\bibitem{Lenz} 
  A.~Lenz,
  %``Lifetimes and heavy quark expansion,''
  Int.\ J.\ Mod.\ Phys.\ A {\bf 30}, no. 10, 1543005 (2015).
	
\bibitem{mv99} 
  M.~B.~Voloshin,
  %``Relations between inclusive decay rates of heavy baryons,''
 in Phys.\ Rept.\  {\bf 320}, 275 (1999).

\bibitem{mv00} 
  M.~B.~Voloshin,
  %``Reducing model dependence of spectator effects in inclusive decays of heavy baryons,''
  Phys.\ Rev.\ D {\bf 61}, 074026 (2000).
\bibitem{aleph1} 
  D.~Buskulic {\it et al.} [ALEPH Collaboration],
  %``A Measurement of the b baryon lifetime,''
  Phys.\ Lett.\ B {\bf 297}, 449 (1992).
		
\bibitem{aleph2} 
  D.~Buskulic {\it et al.} [ALEPH Collaboration],
  %``Measurements of the $b$ baryon lifetime,''
  Phys.\ Lett.\ B {\bf 357}, 685 (1995).
	
\bibitem{aleph3} 
  R.~Barate {\it et al.} [ALEPH Collaboration],
  %``Measurement of the $B$ baryon lifetime and branching fractions in $Z$ decays,''
  Eur.\ Phys.\ J.\ C {\bf 2}, 197 (1998).
	
\bibitem{opal} 
  K.~Ackerstaff {\it et al.} [OPAL Collaboration],
  %``Measurements of the $B_s^0$ and $\Lambda_b^0$ lifetimes,''
  Phys.\ Lett.\ B {\bf 426}, 161 (1998).
	
\bibitem{lhcb1} 
  R.~Aaij {\it et al.}  [LHCb Collaboration],
  %``Precision measurement of the $\Lambda_b^0$ baryon lifetime,''
  Phys.\ Rev.\ Lett.\  {\bf 111}, 102003 (2013).


	
\bibitem{lhcb2} 
  R.~Aaij {\it et al.}  [LHCb Collaboration],
  %``Precision measurement of the ratio of the $\Lambda^0_b$ to $\overline{B}^0$ lifetimes,''
  Phys.\ Lett.\ B {\bf 734}, 122 (2014).
	
\bibitem{pdg} 
  M.~Tanabashi {\it et al.} [Particle Data Group],
  %``Review of Particle Physics,''
  Phys.\ Rev.\ D {\bf 98}, no. 3, 030001 (2018).
	
\bibitem{lhcb19} 
  R.~Aaij {\it et al.} [LHCb Collaboration],
  %``Precision measurement of the $\Lambda_c^+$, $\Xi_c^+$ and $\Xi_c^0$ baryon lifetimes,''
  Phys.\ Rev.\ D {\bf 100}, no. 3, 032001 (2019)
	
\bibitem{mv00-2} 
  M.~B.~Voloshin,
  %``Weak decays Xi(Q) ---> Lambda(Q) pi,''
  Phys.\ Lett.\ B {\bf 476}, 297 (2000).
	
\bibitem{lhcb15}
  R.~Aaij {\it et al.} [LHCb Collaboration],
  %``Evidence for the strangeness-changing weak decay $\Xi_b^-\to\Lambda_b^0\pi^-$,''
  Phys.\ Rev.\ Lett.\  {\bf 115} (2015) no.24,  241801	
	
\bibitem{lhcb19-0} 
  R.~Aaij {\it et al.} [LHCb Collaboration],
  %``Measurement of the mass and production rate of $\Xi_b^-$ baryons,''
  Phys.\ Rev.\ D {\bf 99}, no. 5, 052006 (2019).
	
\bibitem{vs2}
M.~A.~Shifman and M.~B.~Voloshin, Sov.\ J.\ Nucl.\ Phys.\ {\bf 45}, {292} (1987).

\bibitem{Cheng92}
  H.~Y.~Cheng, C.~Y.~Cheung, G.~L.~Lin, Y.~C.~Lin, T.~M.~Yan and H.~L.~Yu,
  %``Heavy flavor conserving nonleptonic weak decays of heavy baryons,''
  Phys.\ Rev.\ D {\bf 46}, 5060 (1992).

\bibitem{sk} 
  S.~Sinha and M.~P.~Khanna,
  %``Beauty-conserving strangeness-changing two-body hadronic decays of beauty baryons,''
  Mod.\ Phys.\ Lett.\ A {\bf 14}, 651 (1999).
	
\bibitem{lv} 
  X.~Li and M.~B.~Voloshin,
  %``Decays $\Xi_b \to \Lambda_{b \pi}$ and diquark correlations in hyperons,''
  Phys.\ Rev.\ D {\bf 90}, no. 3, 033016 (2014).

\bibitem{fm} 
  S.~Faller and T.~Mannel,
  %``Light-Quark Decays in Heavy Hadrons,''
  Phys.\ Lett.\ B {\bf 750}, 653 (2015).

\bibitem{gr1} 
  M.~Gronau and J.~L.~Rosner,
  %``$S$-wave nonleptonic hyperon decays and $\Xi^-_b \to \pi^- \Lambda_b$,''
  Phys.\ Rev.\ D {\bf 93}, no. 3, 034020 (2016).
	
\bibitem{gr2} 
  M.~Gronau and J.~L.~Rosner,
  %``From $\Xi_b \to \Lambda_b \pi$ to $\Xi_c \to \Lambda_c \pi$,''
  Phys.\ Lett.\ B {\bf 757}, 330 (2016)

\bibitem{Cheng15} 
  H.~Y.~Cheng, C.~Y.~Cheung, G.~L.~Lin, Y.~C.~Lin, T.~M.~Yan and H.~L.~Yu,
  %``Heavy-Flavor-Conserving Hadronic Weak Decays of Heavy Baryons,''
  JHEP {\bf 1603}, 028 (2016).
  
	
\end{thebibliography}
\end{document}